\documentclass[journal]{IEEEtran}

\usepackage{graphicx}
\usepackage{amsmath}
\usepackage{tabu}

\begin{document}

\title{A Compact Filter-Bank Waveguide Spectrometer for Millimeter Wavelengths}
\author{Sean Bryan$^*$\thanks{$^*$Email: sean.a.bryan@asu.edu}, George Che, Christopher Groppi, Philip Mauskopf, and Matthew Underhill}

\maketitle

\begin{abstract}
We present the design and measurements of a 90 GHz prototype of a millimeter-wave channelizing spectrometer realized in rectangular waveguide for astronomical instrumentation. The device was fabricated using conventional high-precision metal machining, and the spectrometer can be tiled into a 2D array to fill the focal plane of a telescope. Measurements of the fabricated five-channel device matched well with electromagnetic simulations using HFSS and a cascaded S-matrix approach. This motivated the design of a 54-channel $R$=200 spectrometer that fills the single-moded passband of rectangular waveguide in the 130-175 GHz and 190-250 GHz atmospheric windows for millimeter-wave spectroscopic mapping and multi-object spectroscopy. 
\end{abstract}


\section{Introduction}

The development and optimization of large format bolometric arrays for imaging and polarimetry from ground-based millimeter-wave and sub-millimeter telescopes has helped to revolutionize the fields of cosmology, galaxy evolution and star formation. High resolution imaging and spectroscopy of individual mm-wave sources is now being done by ALMA \cite{hills11}, however spectral surveys over wide sky areas and wide frequency ranges are not practical with ALMA. The next major steps in millimeter-wave imaging and spectroscopy will include several science goals. Large area spectral surveys with moderate spectral resolution (e.g. $R \simeq 50-200$) could be used to characterize large scale structure and the star formation history of the universe using intensity mapping of emission lines such as CO \cite{lidz11} and CII \cite{silva14}. Multi-object mm-wave wide-band spectroscopy with moderate spectral resolution would enable galaxy redshift surveys. Further studies of hot gas in galaxy clusters through the Sunyaev Zeldovich (SZ) effect would be enabled by high angular resolution and moderate spectral resolution instruments. 

One of the main new technologies required to achieve these science goals is an arrayable wideband spectrometer consisting of a compact spectrometer module coupled to highly multiplexable detector arrays. Several groups are working on developing superconducting on-chip spectrometers (e.g. SuperSpec \cite{Bradford, HaileyDunsheath}, Micro-Spec \cite{Cataldo}, DESHIMA \cite{Endo} and CAMELS \cite{Thomas}) based on either filter banks or on-chip gratings. Existing mm-wave spectrometers in the field include imaging Fourier Transform Spectrometers \cite{Naylor} and the Z-Spec waveguide grating-type spectrometer \cite{zspec}. Here we present the design and prototype test results for a compact scalable waveguide filter-bank spectrometer. This spectrometer can be manufactured with standard high precision machining facilities, and is able to be tested both warm and cold independently from the detectors. It can be coupled to simple-to-fabricate highly multiplexable kinetic inductance detectors (KIDs), or to conventional bolometers. The spectrometer is highly complementary to the on-chip spectrometers in several ways. First, it could be naturally used in the currently undeveloped spectral range from 130-250 GHz, which is optimal for measurements of CO line emission and the kinetic SZ effect. Also, this device can be straightforwardly designed to cover a relatively broad spectral resolution compared to the superconducting filters. Finally, instead of using KIDs as the detector technology, the device could alternately be configured as a room temperature wideband backend for mm-wave and cm-wave cryogenic amplifiers without the need for downconverting mixers.

\section{Spectrometer Concept and Measurements of the Prototype}

\begin{figure*}
\begin{center}
\includegraphics[width=0.80\textwidth]{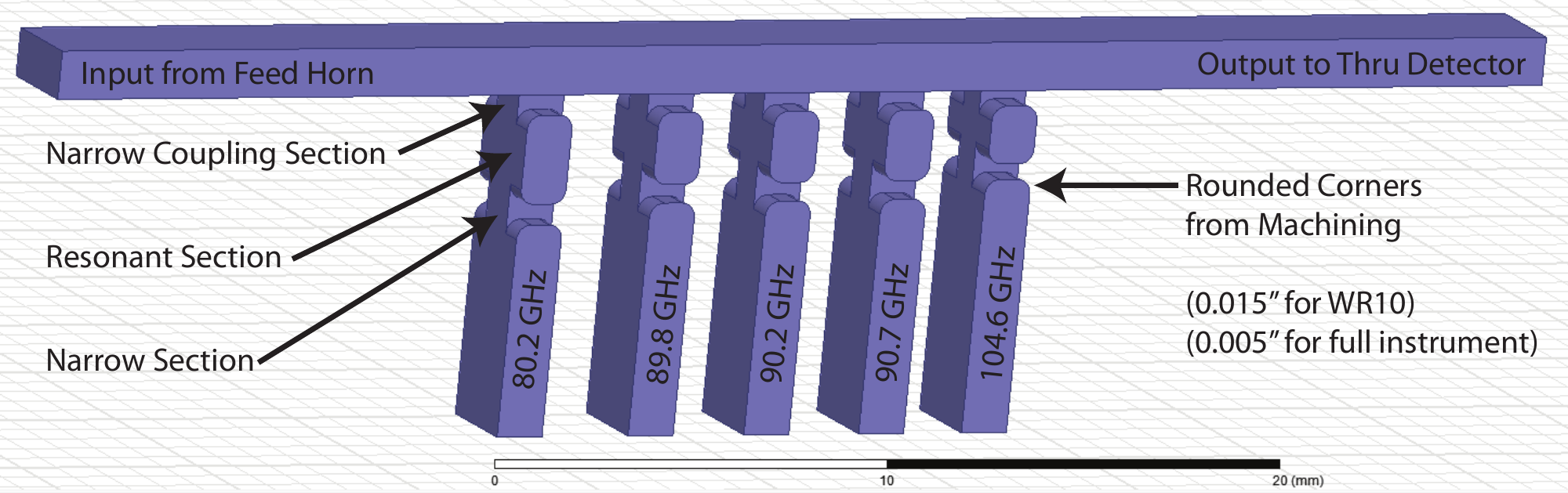}
\caption{Filter concept overview illustrated with the five-channel WR10 prototype. Light from the feed horn (not shown) comes down the main waveguide, and different frequencies are selected off by each of the five channels shown. A device filling the single-moded passband of rectangular waveguide would have 54 spectral channels, and has been designed as shown in Fig. \ref{passbands}. The rounded corners are from the machining process, and are treated in the simulation. \label{filter_concept_overview}}
\end{center}
\end{figure*}

The design concept of the filter-bank spectrometer is illustrated in Fig.~\ref{filter_concept_overview}, which shows a drawing of the five-channel prototype filter bank we have fabricated and tested. A direct-drilled circular feed horn ~\cite{tan2011a} followed by a circular to rectangular transition (not shown) couples light from the sky onto the main rectangular waveguide. Each channel connects to the main waveguide through an E-plane tee that uses an evanescent coupling section into a half-wavelength resonating cavity. An identical narrow section on the other end of the resonant cavity defines the resonating length. The radiation then terminates on a detector. The narrow sections of waveguide have a cutoff frequency that is 50$\%$ higher than the center frequency of the passband of the channel. Because the light from the sky is below the cutoff frequency of the narrow section, this means the section's impedance is effectively capacitive. However, on-resonance the cavity section is effectively inductive, which tunes out the capacitive sections, allowing the radiation to pass through to that channel's detector. At frequencies far from resonance, this cancellation does not take place, and light does not couple to the detector. The center frequency of each channel is tuned by adjusting the length of the resonant section, and the bandwidth is defined by adjusting the length of the narrow capacitive sections. Because the operation of the device depends on the narrow coupling sections having cutoff frequencies above the maximum frequency incident on the device, and because all of our modeling is in the single-mode limit, the light from the sky will need to be filtered to keep the device operating in the single mode limit and prevent spurious coupling of high frequencies down the spectrometer channels.

\begin{figure}
\begin{center}
\includegraphics[width=0.48\textwidth]{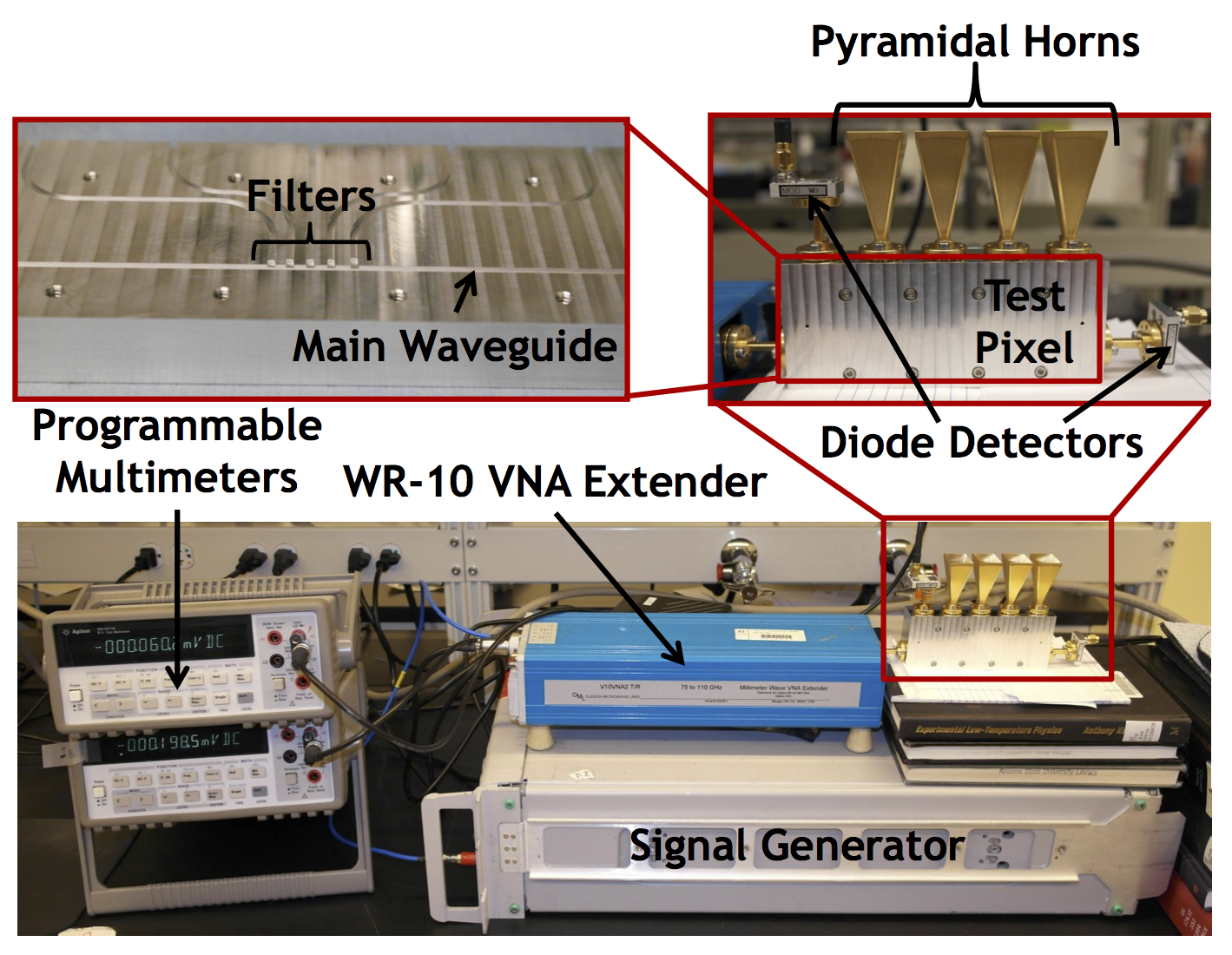}
\caption{Photograph showing the five-channel test device and the measurement setup. The source component of a VNA extender was used to generate millimeter waves at the input port. The outputs of each spectrometer channel were either terminated with a feed horn and absorbing AN-72, or measured with a diode detector. This yielded a measurement of the optical efficiency and bandwidth of each spectrometer channel. \label{measurement_photo}}
\end{center}
\end{figure}

To verify this concept, we constructed and tested a prototype five-channel spectrometer machined from aluminum for the WR10 band. This band was chosen to allow testing using a WR10 VNA extender. The dimensions were chosen based on simulations in HFSS. Because this spectrometer is realized in waveguide, it can operate at both room and cryogenic temperatures, which enabled simple and rapid testing. The prototype device has one channel with a center frequency $f_c$ at 80 GHz, three closely spaced channels near 90 GHz, and a fifth channel at 105 GHz, each with $R \equiv f_c / \Delta f \sim 200$ resolution. Just as would be possible with a larger multichannel spectrometer, we used E-plane split block construction with conventional alignment pins. Nominally no RF current flows across the split. This prototype was machined on a 5-micron tolerance CNC milling machine. Tolerances of 1-2 micron tolerances that would improve operation at 200-300 GHz are regularly achieved on standard high-precision CNC milling machines.

\begin{figure}
\begin{center}

\includegraphics[width=0.48\textwidth]{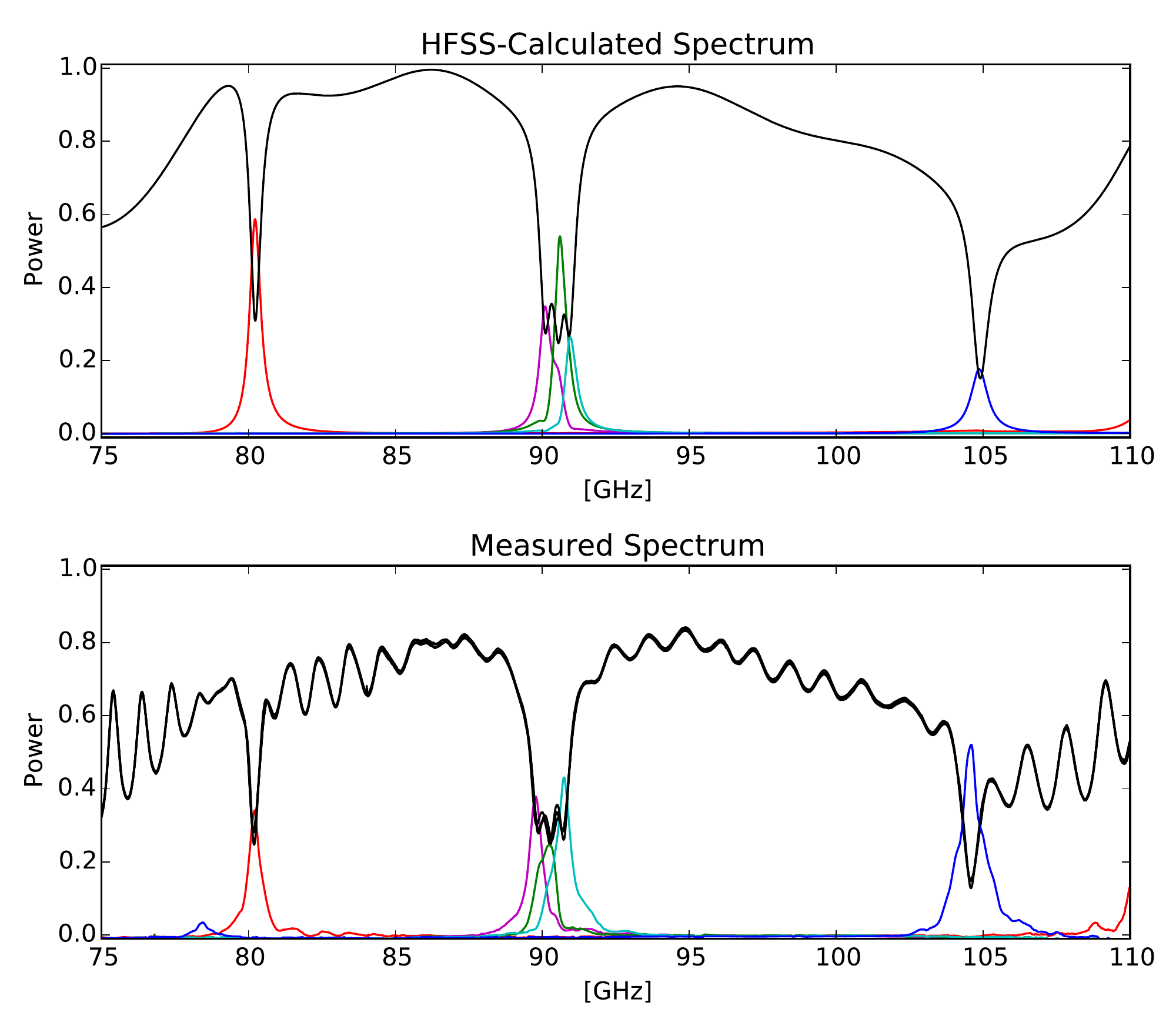}

\footnotesize
{\tabulinesep=3pt
   \begin{tabu} {|c|c|c|c|c|c|}
    Measured & Calculated &   Meas. & Calc. &   Meas. & Calc. \\
    Frequency & Frequency &   $R$ & $R$ &   OE & OE \\ \hline \hline
    \phantom{1}80.20 GHz          & \phantom{1}80.22 GHz            &   178                 & 182       &  34$\%$ & 59$\%$            \\ \hline
    \phantom{1}89.78 GHz          & \phantom{1}90.09 GHz            &   180                 & 120       &  38$\%$ & 35$\%$            \\ \hline
    \phantom{1}90.23 GHz          & \phantom{1}90.60 GHz            &   120                 & 193       &  24$\%$ & 54$\%$            \\ \hline
    \phantom{1}90.75 GHz          & \phantom{1}90.95 GHz            &   158                 & 207       &  43$\%$ & 26$\%$            \\ \hline
    104.60 GHz         & 104.87 GHz           &  135                 & 154                 &  52$\%$ & 18$\%$  \\ \hline
\end{tabu}}
\caption{Measured and HFSS-calculated performance of the five-channel WR10 test device. The top panel shows the HFSS-simulated passbands of the individual channels in the prototype device, shown in colors ranging from red to blue. The simulated coupling of the thru detector is shown in black. Measured performance is shown in the middle panel, and the table at the bottom compares the measured and simulated center frequencies, spectral resolutions $R$, and optical efficiencies (OE). \label{test_part_linear}}
\end{center}
\end{figure}

We measured the passbands of each channel by terminating all but one channel using standard gain horns (Quinstar QWH-WPRS00) to dump the power onto absorbing AN-72 foam. We then used a broadband square-law diode detector (Pacific Millimeter Products WD) to measure the passband of the remaining channel, and put a second detector on the thru port. We generated millimeter waves using the source component of an OML V10VNA2 VNA extender driven by an Anritsu 69247A microwave frequency generator. The VNA extender produced enough millimeter-wave power to cause the detectors to have a non-linear response, so a 20 dB attenuator was mounted at the VNA extender source port to keep the detectors in their linear response power range.

The passband of each spectrometer channel was measured by sweeping the frequency at the signal generator to generate power in 0.025 GHz steps from 75 to 110 GHz, and recording the DC voltage on the detector at each step. A detector was mounted to each spectrometer channel in succession, while terminating all the other channels with the horns. The gain variation of the each detector across the passband, as well as an absolute power calibration, was established by performing a frequency sweep with each detector attached directly to the source (with the attenuator still in place). Dividing the spectra taken with the detectors attached to the spectrometer by these reference spectra gives a measurement of both the absolute optical efficiency and the shape of the passband.

To guide the design of the initial prototype, and for comparison with the measurements, we simulated the entire five-channel structure in HFSS. The results of the measurements and the simulation are shown in Fig.~\ref{test_part_linear}. The measured optical efficiency of each channel is high, broadly consistent with the modeling. The measured center frequencies agree with the calculation to better than $0.5\%$, and the measured spectral resolutions agree to within $30\%$. Out of band coupling is observed at the -25 dB level. The standing wave pattern in the thru detector suggests an optical path length of roughly 30 cm, which could suggest a standing wave between imperfect terminations in the horns or detectors, and another imperfect match somewhere inside the VNA extender. Overall, there is good qualitative agreement between the measurement and the HFSS simulation, and the selectivities and optical efficiencies of the channels are good, suggesting that we can use HFSS to design a spectrometer with more channels.

\section{Cascaded HFSS Simulation Method}

\begin{figure*}
\begin{center}
\includegraphics[width=1.0\textwidth]{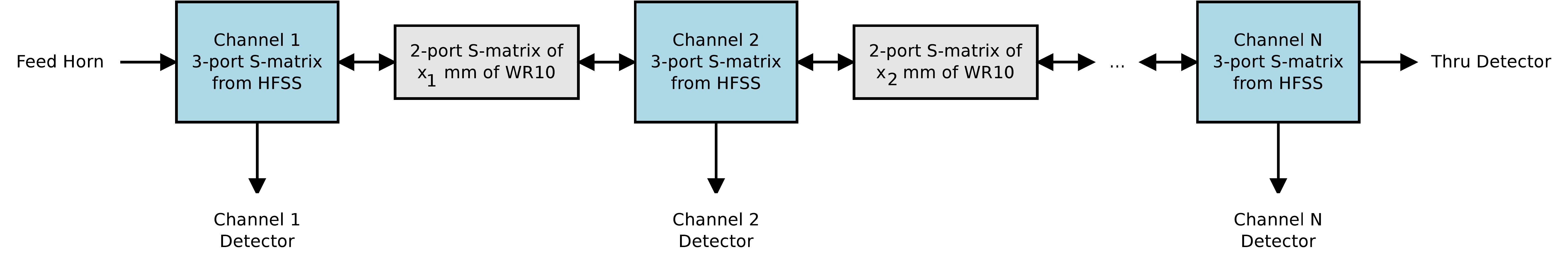}
\caption{Flowchart illustrating the cascaded S-matrix simulation method for the spectrometer. The blue boxes represent the three-port S-matrices of single spectrometer channels simulated using HFSS. The light grey boxes represent the two-port S-matrices of rectangular waveguide that connect the individual channels, which are calculated analytically. The S-matrices are cascaded using the scikit-rf module in python. This method successfully simulates the five-channel prototype, and is far faster in simulating a multi-channel device than HFSS would be. \label{cascade_flowchart}}
\end{center}
\end{figure*}

\begin{figure}
\begin{center}
\includegraphics[width=0.48\textwidth]{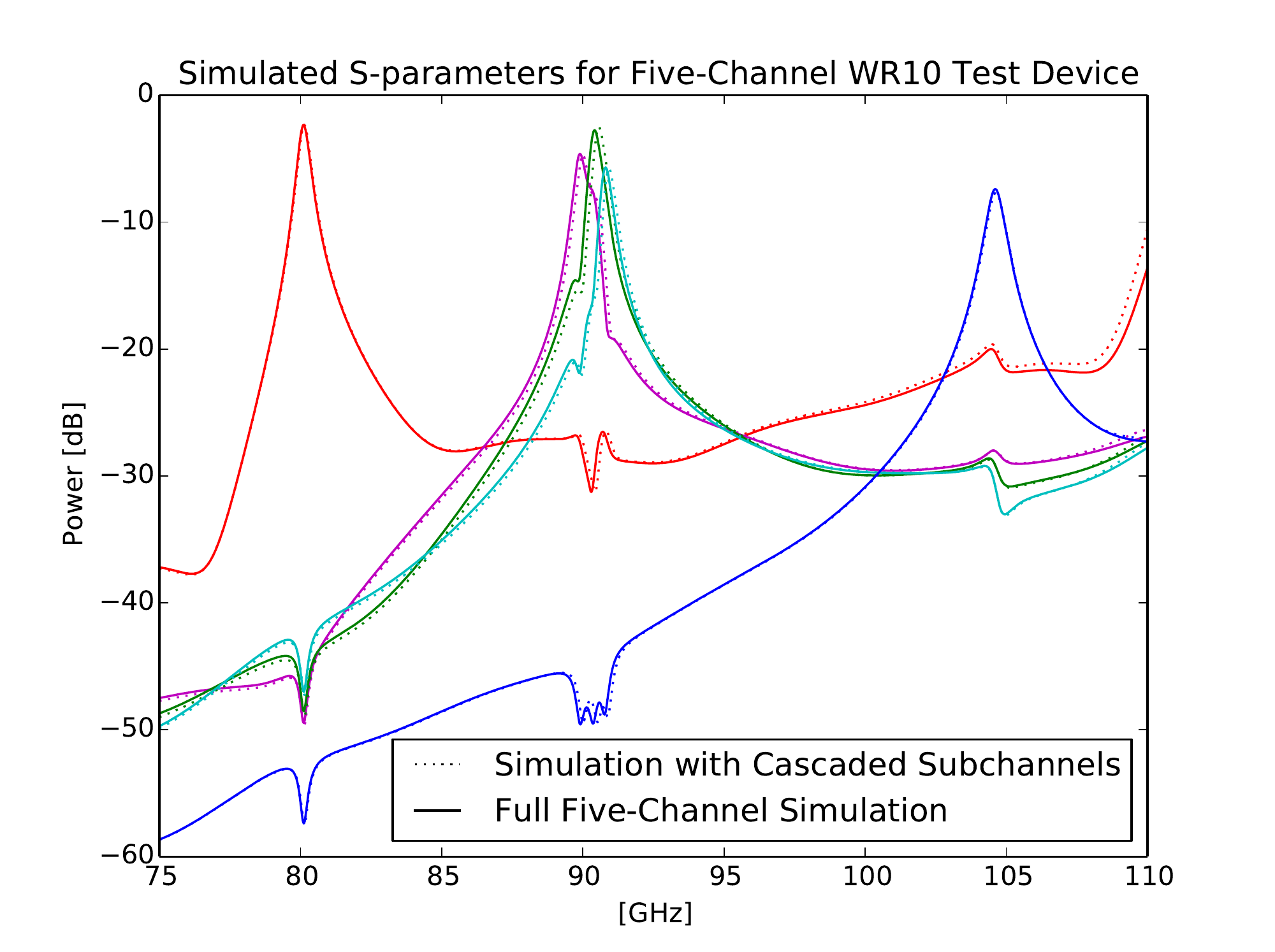}
\caption{Passbands for the five-channel device simulated with HFSS alone, and a cascaded HFSS single-channel simulation for comparison. Since the cascade simulation reproduces the details of the HFSS simulation, down to -60 dB, this motivates using the method to design a multi-channel device that would be too large to simulate with HFSS alone. \label{comparison}}
\end{center}
\end{figure}

Having verified the spectrometer concept, we designed a spectrometer that has a large number of channels spread across a wide passband. Here we consider two bands, 130-175 GHz and 190-250 GHz. At a spectral resolution of $R$=200, 54 spectrometer channels are needed to fill the single-moded passband of rectangular waveguide, assuming we place the center frequencies of each channel at the half-power points of each of its nearest neighbors. A full 54-channel structure would be so large that it would require too much RAM and too much CPU time to reasonably simulate in a single HFSS run on a workstation. However, we found that cascading the S-matrices of single-channel HFSS simulations was as accurate as a full simulation, and can be done for a device with a large number of channels. We verified the accuracy of this cascade method by using it to simulate the five-channel prototype device, and comparing that simulation with the full HFSS simulation of the entire five-channel structure.

An overview of the cascade method is shown in Fig.~\ref{cascade_flowchart}. We start by drawing the structure of an individual spectrometer channel, and simulating it in HFSS. The full 3D simulation is necessary to calculate the effects of the rounded corners left by the machining process, and the fringing fields around the waveguide tee. These simulations yield the S-matrix of an individual channel. Since the field distribution at each of the three ports of an individual channel is nearly identical to the fundamental mode of rectangular waveguide, the S-matrices of the individual channels can be cascaded together, both to each other and to sections of rectangular waveguide, to yield an accurate model of the total S-matrix of the entire system. We used a function in the scikit-rf module in python to perform the cascade, which is an implementation the subnetwork growth algorithm \cite{compton89}. The S-matrix of a section of rectangular waveguide is composed of simple analytic functions, which makes the cascade simulation fast. Cascading is much more accurate than naively multiplying the individual transmission and reflection values, since it properly treats multiple reflections among the internal sub-structures in the system.

A comparison of the simulated spectrometer passbands using both methods for the five-channel device is shown in Fig.~\ref{comparison}. We found that we could reproduce all the details of the five-channel full simulation down to the -60 dB level by using the cascade method. Using this equivalent method for the 54-channel structure makes it possible to simulate and design the device needed for the full instrument. Simulating an entire 54-channel structure with the cascade method takes only 3 hours of workstation CPU time to generate the single-channel HFSS simulations, and 2 minutes to cascade them together.

\section{Simulation Results for a 54 Channel Device}

Each channel in the spectrometer has three dimensions: the length of the resonator section, the length of the coupling sections, and the width of the coupling section. For a 54-channel device, that means there are 162 free parameters that needed to be adjusted to achieve the target center frequency and $R=200$ resolution for each channel. Tuning all 162 parameters individually by hand is not feasible, so instead we determined the optimal parameters by interpolating between hand-tuned designs. Here we illustrate this process for selecting the resonator length dimensions, the coupling section lengths and widths were chosen in an identical fashion. First, we hand-tuned the dimensions of six channels spread evenly across the desired passband by interactively running HFSS simulations. The set of center frequencies $f_{design}$ of the hand-tuned channels was
\begin{align}
f_{design}^i = \{&80.00~\mathrm{GHz},85.42~\mathrm{GHz},89.79~\mathrm{GHz}, \dots , \nonumber \\ 
&95.18~\mathrm{GHz},100.46~\mathrm{GHz},105.26~\mathrm{GHz} \}
\end{align}
and the resonator lengths corresponding to those center frequencies determined by hand-tuning were
\begin{align}
l_{design}^i = \{ &2.355~\mathrm{mm},2.068~\mathrm{mm},1.905~\mathrm{mm},  \nonumber \\ 
&1.739~\mathrm{mm},1.617~\mathrm{mm},1.523~\mathrm{mm} \}.
\end{align}
The coupling section widths were set such that their cutoff frequency is 50$\%$ higher than the center frequency, and the coupling section lengths were adjusted to yield the desired $R$=200 resolution. These sets of hand-tuned dimensions can be thought of as a ``dataset'' of designs with known center frequencies and the desired resolution.

We then scaled these designs to center frequencies across the entire band. Putting center frequencies across the whole passband and at the half-power points of their nearest neighbors yielded a set of 54 desired center frequencies
\begin{align}
f_{NChn}^i = \{ &80.00~\mathrm{GHz},80.41~\mathrm{GHz},80.82~\mathrm{GHz}, \dots , \nonumber \\ 
&103.93~\mathrm{GHz},104.46~\mathrm{GHz},105.00~\mathrm{GHz} \}.
\end{align}
This can be thought of as Nyquist-sampling the spectral band. The hand-tuned designs needed to be scaled somehow to determine the resonator lengths to yield these desired center frequencies. Since all channels will use the same WR10 waveguide, the designs were scaled by the in-guide wavelength corresponding to each center frequency. For rectangular waveguide in the fundamental mode, the in-guide wavelength $\lambda_g$ for a frequency $f$ is
\begin{equation}
\lambda_g (f) = \frac{c/f}{\sqrt{1 - \Big(\frac{c/f}{2a}\Big)^2}},
\end{equation}
where $c$ is the speed of light, and $a$ is the width of the waveguide. To generate the resonator lengths $l_{NChn}^i$ corresponding to the desired center frequencies, we linearly interpolated between the hand-tuned channels, using the guide wavelength. The lengths are therefore
\begin{equation}
l_{NChn}^i = \mathrm{interp}(\lambda_g(f_{design}^i),l_{design}^i,\lambda_g(f_{NChn}^i)),
\end{equation}
where the function $\mathrm{interp}(x_{data},y_{data},x)$ linearly interpolates between the datapoints $(x_{data},y_{data})$ to estimate the $y$ value corresponding to the input $x$. The dimensions for the coupling section widths and lengths were interpolated similarly from the hand-tuned designs.

Once we had those dimensions, we ran HFSS simulations for each single channel. This process was automated using the Matlab API for HFSS, and the individual S-matrices for each individual channel were stored to disk. This method yielded center frequencies that were fairly close to the desired values. We took the results of this simulation run as ``data'' and interpolated between them to get designs which are even closer to the desired center frequencies. This process, and slightly tweaking a few of the channels by hand, yielded a set of dimensions where the center frequencies of the final 54-channel cascaded simulation matched the design goal to within an RMS of $0.05\%$ and the spectral resolutions were within an RMS of $25\%$ of $R=200$. 

To determine the optimal physical spacing between channels along the main waveguide, we cascaded the 54 individual channel simulations together to yield the 56 port S-matrix that simulates the performance of all of the channels in the full device. Since changing the spacings only requires repeating the cascade step in the simulation, not resimulating in HFSS, changing the spacings only takes 2 minutes of workstation CPU time to recompute. We were therefore able to simulate many spacings and choose the best one. We chose to arrange the channels from low to high frequency, and have a constant spacing between each channel. Resimulating over a range of spacings showed that the optimal channel spacing is 3.065 mm, which is 3/4 of a wavelength in WR10 guide at 94.2 GHz. Initial simulations using a different spacing between each pair of channels did not yield better performance than using constant spacing, but we are investigating this more to see if further performance improvements are possible.

\begin{figure}
\begin{center}

\includegraphics[width=0.48\textwidth]{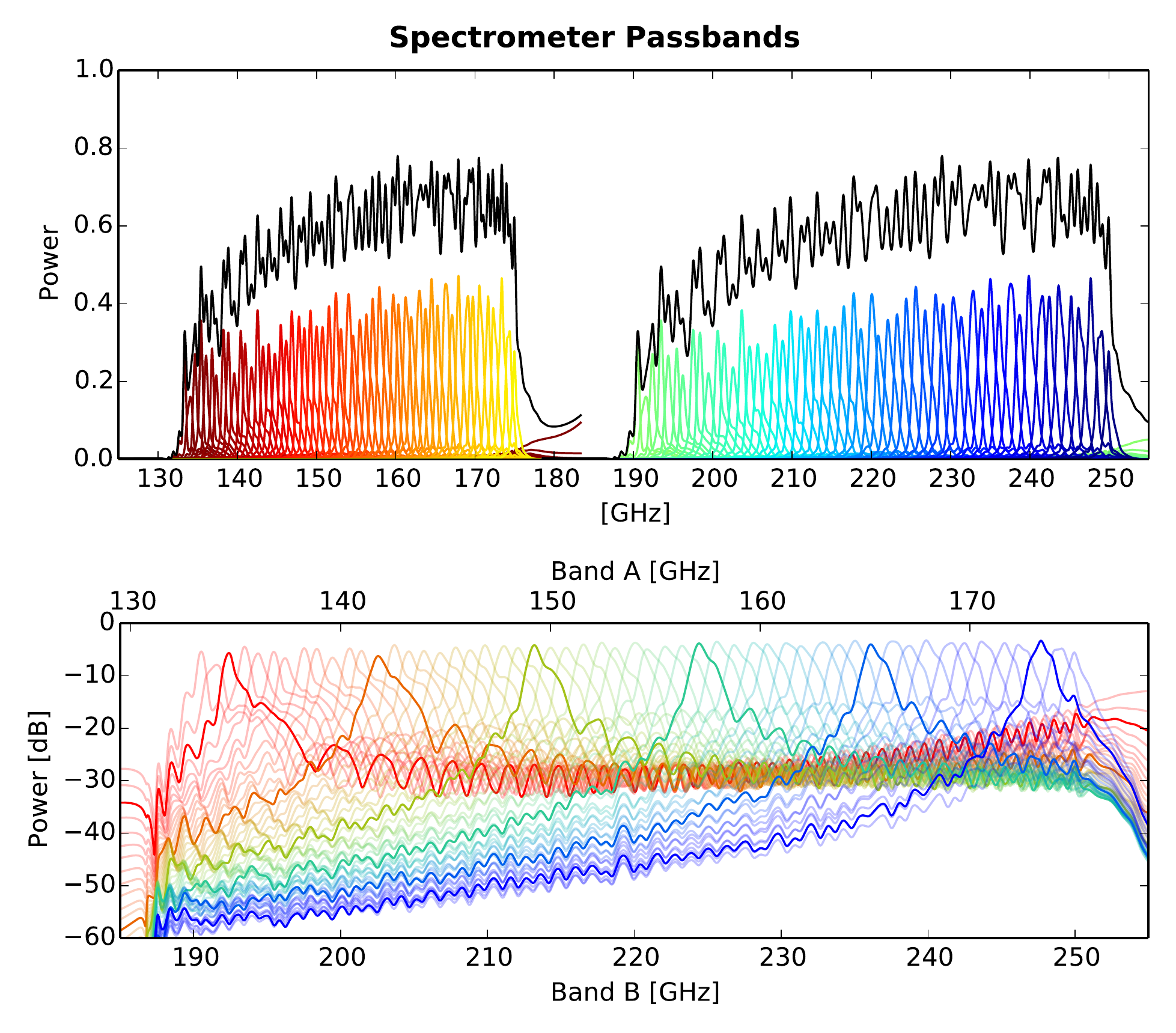}

\caption{Simulated passbands of a 54-channel spectrometer. The top panel is on a linear scale and shows the 54 Band A passbands which cover 135-170 GHz and the 54 Band B passbands which cover 190-245 GHz. The black curve shows the sum total of all the passbands. The optical efficiency of the individual channels ranges from roughly 0.25 to 0.4, which compares favorably to the idea-impedance-match case of 0.5. The bottom panel shows the simulation down to -60 dB, with selected channels highlighted. Band A is indicated on the top x-axis, and Band B is on the bottom x-axis. Out of band coupling is simulated to be at the -20 to -30 dB level, or better.\label{passbands}}
\end{center}
\end{figure}

The WR10 design was simulated without any conductor loss in the model. The good agreement between the lossless simulations and the measurements of the WR10 prototype suggest that in principle we can scale the dimensions and the simulation to other slightly higher frequency passbands. Two bands lying in atmospheric windows that are interesting for observing the SZ effect and CO/CII spectroscopy are the 130-175 GHz band, and a 190-250 GHz band. The gap between the two bands is to avoid a strong atmospheric absorption line. The simulated passbands of the individual channels of both a 130-175 GHz band device and a 190-250 GHz device are shown in Fig.~\ref{passbands}. The passbands of neighboring channels cross at the half-power point, which Nyquist samples the entire bandwidth. The entire 130 GHz to 250 GHz passband does not fit in the single-moded bandwidth of a rectangular waveguide, so we will use either a single horn with a diplexer, or two independent feed horns, to place the full bandwidth into a lower Band A and an upper Band B, on either side of the atmospheric line at 180-185 GHz. Since the device is small enough to fit under the footprint of the feed horn, a linear array of these spatial pixels is formed in one direction, which all feed a single wafer of KID detectors. These linear arrays can then be tiled in the other direction, like vertical cards in a motherboard, to form a filled focal plane array of spectrometers.

\begin{figure}
\begin{center}
\includegraphics[width=0.48\textwidth]{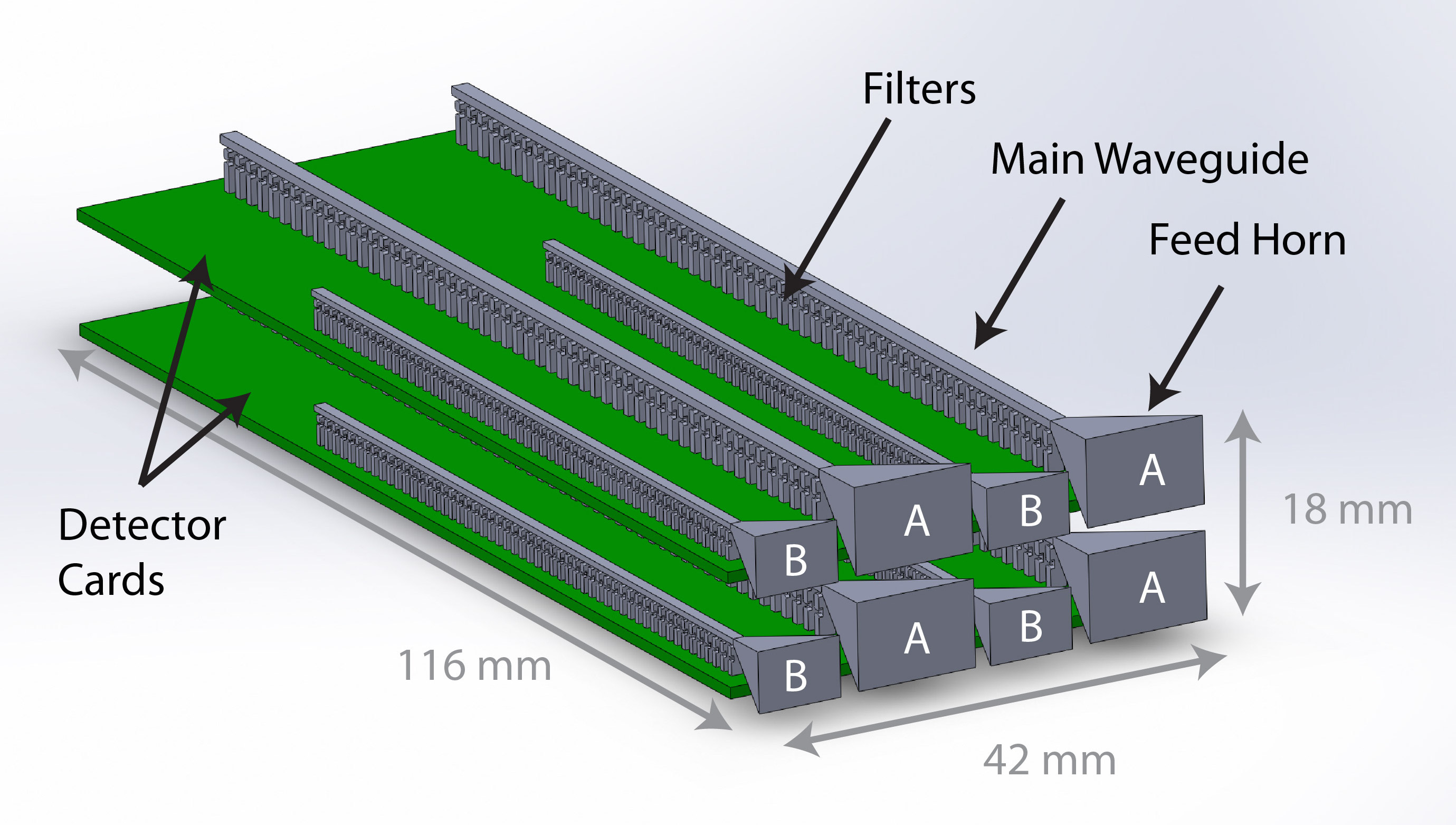}
\caption{Illustration of a filled focal plane concept, drawn  in 3D to scale. Light from the telescope comes in from the bottom-right of the illustration, couples onto the f/3 feed horns and into the main waveguides, and is selected out into frequency channels by the filter bank. KID detectors are in the cards shown in green. In this drawing, Band A (the frequencies below the atmospheric line at 185 GHz) and Band B (the channels above the 185 GHz line) are fed by separate feed horns, but each pixel still only takes up about 10x20 mm of focal plane area. We are investigating a diplexer to possibly enable further miniaturization, and to keep the system operating in the single-mode limit, by feeding both bands with a single smaller horn.\label{focal_plane}}
\end{center}
\end{figure}

A concept schematic of a $2 \times 2$ array of spatial pixels is shown in Fig.~\ref{focal_plane}. For a single spatial pixel in the focal plane, the light will come in from the horn, down the main waveguide, and be directed into the 54 individual spectral channels. In the prototype WR10 design, the physical center-to-center spacing between channels is a constant 3.065 mm. Scaling the design to Band A gives a total device length of 96 mm, and Band B will have a length of 68 mm. These dimensions are small enough that fabricating the corresponding detector array cards in standard cleanroom processes will be feasible.

\subsection{Loss and Machining Tolerances}

Scaling the lossless simulation to design higher frequency devices is a valid approximation as long as conductor loss continues to be negligible in the higher passbands. For a rectangular waveguide, the attenuation constant due to conductor loss is
\begin{equation}
\alpha_c = \sqrt{\frac{\pi f \mu_0}{\sigma}} \frac{b + 2 a^3 \Big( \frac{f}{c}\Big)^2}{a^3 b \eta \Big( \frac{f}{c}\Big) \sqrt{\Big( \frac{2f}{c}\Big)^2 - \Big( \frac{1}{a}\Big)^2}},
\end{equation}
where $a$ and $b$ are the width and height of the waveguide, $\sigma$ is the metal conductivity, and $\eta$ is the free space impedance \cite{pozar12}. Since the resonator quality factor is defined in terms of fractional energy lost per cycle, the $Q$ due to conductor loss is approximately
\begin{equation}
Q_{loss} = \frac{2 \pi}{1 - e^{-\alpha_c \lambda_g}}.
\end{equation}
This loss $Q$ will degrade the actual spectral resolution $R_{tot}$ from its nominal lossless design value $R$ by
\begin{equation}
\frac{1}{R_{tot}} = \frac{1}{R} + \frac{1}{Q_{loss}}.
\end{equation}

For the WR10 prototype with aluminum's room temperature DC conductivity \cite{ekin07}, the calculated attenuation constant of aluminum WR10 waveguide at 105 GHz is 0.25 $\textrm{m}^{-1}$. This implies a calculated $Q_{loss}$ of approximately 7,000, which is high enough to explain the good performance of our prototype. Since our prototype was roughly $R=200$, directly measuring the impact of this loss $Q$ would have required fabricating and testing another test device with high design $R$, which we have not yet done.

Scaling up the maximum operating frequency, and the corresponding reduction in the waveguide dimensions, both increase the loss. However, for a maximum operating frequency of 250 GHz, the loss Q is calculated to be roughly 4,500, which is still high enough that it would not be expected to limit the performance of a $R=200$ device. Sputter coating a superconducting niobium layer onto the device would eliminate conductor loss below $\sim700$ GHz.

\begin{figure}
\begin{center}
\includegraphics[width=0.48\textwidth]{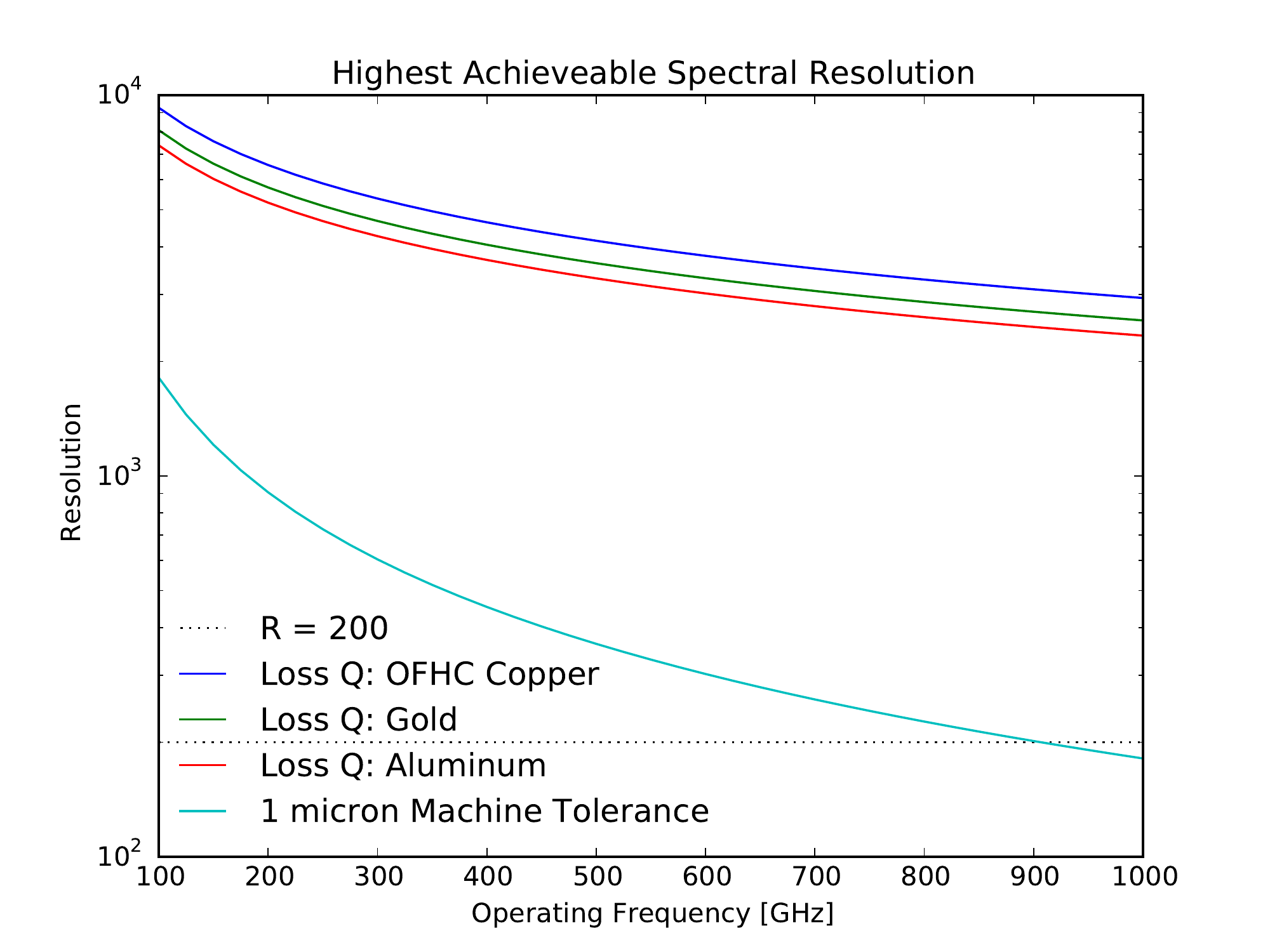}
\caption{Calculation of the limiting spectral resolution due to conductor loss and machining tolerance. The top three curves show loss Q calculated with literature room temperature conductivity values for OFHC copper, gold, and aluminum, which range from roughly 2,000 to 8,000. The bottom curve shows the highest resolution for which nearby channels could be distinguished with a 1 micron tolerance milling machine. This shows that a $R$=200 spectrometer would not be limited by either effect at operating frequencies below roughly 700 GHz. \label{loss_Q}}
\end{center}
\end{figure}

Another factor that will limit the performance of the spectrometer is machining tolerance. If the dimensions of nearby channels in the spectrometer differ by less than the mechanical tolerance to which they are fabricated, then those channels cannot be resolved. This means that the maximum practical resolution of our half-wave resonator channels is approximately
\begin{equation}
R_{max} = \frac{\lambda_g / 2}{\mathrm{Machine~Tolerance}}.
\end{equation}
At ASU, and at other high-precision machine shops, we have a precision milling machine that regularly achieves a 1 micron tolerance, which means that as shown in Fig.~\ref{loss_Q}, a $R=200$ device should not be limited by machining tolerance below about 700 GHz.

\section{Conclusions}

Compact spectrometers that can be tiled into focal plane arrays are an important enabling technology for the next generation of millimeter-wave and sub-millimeter astronomy. Complimentary to the other technologies currently under development, the measurements and modeling of the waveguide filter-bank spectrometer presented here show that it is a promising approach for future instruments. Measurements of the prototype show that cascading the S-matrices from HFSS simulations of individual channels is a good way to model this class of device. This enabled us to design a full $R$=200 spectrometer that fills single-moded passband of rectangular waveguide. We are currently scaling up our prototype to a five-channel test device for the WR5 band (160-210 GHz). We will then test it with another VNA extender at room temperature to verify that conductor loss and machining tolerances do not limit the performance at higher frequencies. We then plan to fabricate and test a full 54-channel device with cryogenic detectors for the 130-175 GHz and 190-250 GHz bands to verify that the optical efficiency and spectral performance are as good as the modeling predicts.

\bibliographystyle{unsrt}
\bibliography{bibliography}

\begin{IEEEbiography}[{\includegraphics[width=1in,height=1.25in,clip,keepaspectratio]{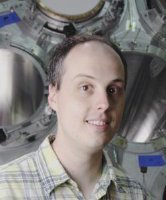}}]{Sean Bryan}
received the PhD degree from Case Western Reserve University, and is currently a postdoctoral researcher at Arizona State University. For his PhD, he worked on the Spider telescope array to measure the Cosmic Microwave Background, which flew successfully on a high-altitude balloon flight from Antarctica in the 2014-2015 season. At Arizona State, he is working on developing Kinetic Inductance Detectors and feed structures for use in millimeter and sub-millimeter wavelengths in astronomy.
\end{IEEEbiography}

\begin{IEEEbiography}[{\includegraphics[width=1in,height=1.25in,clip,keepaspectratio]{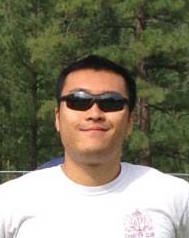}}]{George Che}
received the A.B. degree in physics from Princeton University in 2012, and is currently a PhD candidate in the School of Earth and Space Exploration at Arizona State University. His research interests are in millimeter and sub-millimeter wavelength astronomical instrumentation, and science education. He has designed microwave feed structures and Kinetic Inductance Detectors, fabricates devices in a cleanroom facility, and is a collaborator on several multi-institution telescope teams.
\end{IEEEbiography} 

\begin{IEEEbiography}[{\includegraphics[width=1in,height=1.25in,clip,keepaspectratio]{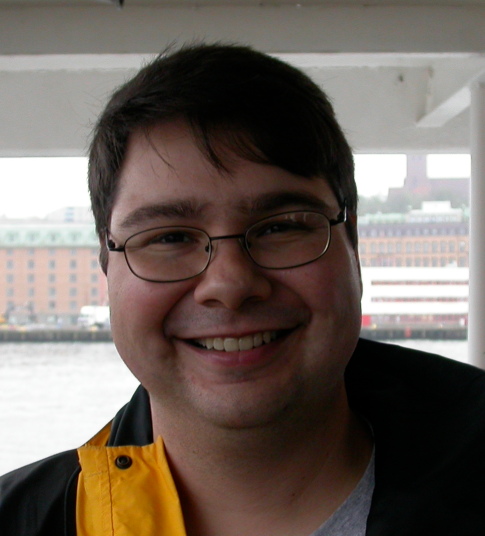}}]{Christopher Groppi}
received the PhD degree from the University of Arizona, and is an assistant professor at Arizona State University. He is an experimental astrophysicist interested in the process of star and planet formation and the evolution and structure of the interstellar medium. His current research focuses on the design and construction of state of the art terahertz receiver systems optimized to detect the light emitted by molecules and atoms in molecular clouds, the birthplace of stars. Development of multi-pixel imaging arrays of terahertz spectrometers is a key technology for the advancement of astrophysics in this wavelength regime. He is participating in several research efforts to develop advanced terahertz imaging arrays for ground based and suborbital telescopes. He also applies terahertz technology developed for astrophysics to a wide range of other applications including Earth and planetary science remote sensing, hazardous materials detection and applied physics.\end{IEEEbiography} 

\begin{IEEEbiography}[{\includegraphics[width=1in,height=1.25in,clip,keepaspectratio]{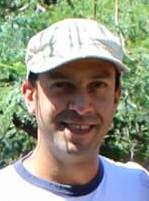}}]{Philip Mauskopf}
received the PhD degree from the University of California, Berkeley, and is a professor at Arizona State University. His background is primarily in experimental cosmology - in particular designing and building new types of instruments for measuring signals from the most distant objects in the universe. His other interests include solid state physics, atmospheric science and quantum communications and cryoptography. He particularly enjoys the opportunity to pursue interdisciplinary projects within the ASU community as well as collaborating with researchers at other universities and research institutions. Before starting at ASU in 2012, he was a Professor of Experimental Astrophysics at Cardiff University in the UK since 2000 where he helped to start a world-leading group in the area of astronomical instrumentation for terahertz frequencies.\end{IEEEbiography}

\begin{IEEEbiography}[{\includegraphics[width=1in,height=1.25in,clip,keepaspectratio]{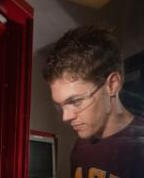}}]{Matthew Underhill}
received the BS degree from Arizona State University, and is currently a machinist there. He specializes in high precision machining and mechanical design for microwave astronomy. He has fabricated feed horns, optical elements, gratings, and other high precision components for millimeter and sub-millimeter wavelengths.
\end{IEEEbiography}

\end{document}